\documentclass[12pt, titlepage, reqno]{article}
\usepackage[authoryear, round]{natbib}
\usepackage{amssymb, latexsym}
\usepackage{amsmath}
\usepackage{amsfonts}
\usepackage{graphicx, centernot}
\usepackage{amsthm}
\usepackage{setspace}
\usepackage{booktabs}
\usepackage{bm}
\usepackage{enumerate}
\usepackage[margin = 1in]{geometry}
\usepackage{titlesec}
\usepackage{fixfoot}
\usepackage{xspace}

\DeclareFixedFootnote{\eqc}{Equal contribution}
\titleformat*{\section}{\normalsize\bfseries}
\titleformat*{\subsection}{\normalsize\bfseries}

\theoremstyle{plain}
\newcommand{\Pro}{\text{P}}
\newcommand{\E}{\mathbb{E}}
\newcommand{\bell}{\boldsymbol{\ell}}
\newcommand{\ci}{\mathrel{\text{\scalebox{1.07}{$\perp\mkern-10mu\perp$}}}}

\newcommand{\balpha}{\boldsymbol{\alpha}}
\newcommand{\bbeta}{\boldsymbol{\beta}}
\newcommand{\boeta}{\boldsymbol{\eta}}
\newcommand{\bgamma}{\boldsymbol{\gamma}}
\newcommand{\beps}{\boldsymbol{\epsilon}}
\newcommand{\btheta}{\boldsymbol{\theta}}
\newcommand{\bSigma}{\boldsymbol{\Sigma}}

\newcommand{\bO}{\boldsymbol{O}}
\newcommand{\bA}{\boldsymbol{A}}
\newcommand{\bU}{\boldsymbol{U}}
\newcommand{\bZ}{\boldsymbol{Z}}
\newcommand{\bW}{\boldsymbol{W}}
\newcommand{\bX}{\boldsymbol{X}}
\newcommand{\bY}{\boldsymbol{Y}}

\newcommand{\bz}{\boldsymbol{z}}

\newcommand{\Var}{\text{Var}}

\begin{document}

\vspace*{0.05in}

\thispagestyle{empty}

\begin{center}

\begin{spacing}{2.0}
{\Large \textbf{Semi-parametric estimation of biomarker age trends with endogenous medication use in longitudinal data}} \\
\end{spacing}

\vspace{4ex}

Andrew J. Spieker$^{1}$, Joseph A.C. Delaney$^{2}$, and Robyn L. McClelland$^{3}$\\
 
\vspace{5mm}

$^{1}$ Department of Biostatistics, Vanderbilt University Medical Center \\
$^{2}$ College of Pharmacy, University of Manitoba \\
$^{3}$ Department of Biostatistics, University of Washington \\

\end{center}

\clearpage

\vspace*{1.2in}

\begin{center}
\textbf{Abstract}
\end{center}

\begin{spacing}{1.0}
In cohort studies, non-random medication use can pose barriers to estimation of the natural history trend in a mean biomarker value---namely, the association between a predictor of interest and a biomarker outcome that would be observed in the total absence of biomarker-specific treatment. Common causes of treatment and outcomes are often unmeasured, obscuring our ability to easily account for medication use with assumptions commonly invoked in causal inference such as conditional ignorability. Further, without a high degree of confidence in the availability of a variable satisfying the exclusion restriction, use of instrumental variable approaches may be difficult to justify. Heckman’s hybrid model with structural shift (sometimes referred to less specifically as the treatment effects model) can be used to correct endogeneity bias via a homogeneity assumption (i.e., that average treatment effects do not vary across covariates) and parametric specification of a joint model for the outcome and treatment. In recent work, we relaxed the homogeneity assumption by allowing observed covariates to serve as treatment effect modifiers. While this method has been shown to be reasonably robust in settings of cross-sectional data, application of this methodology to settings of longitudinal data remains unexplored. We demonstrate how the assumptions of the treatment effects model can be extended to accommodate clustered data arising from longitudinal studies. Our proposed approach is semi-parametric in nature in that valid inference can be obtained without the need to specify any component of the longitudinal correlation structure. As an illustrative example, we use data from the Multi-Ethnic Study of Atherosclerosis to evaluate trends in low-density lipoprotein by age and gender. Results from a collection of simulation studies, as well as our illustrative example, confirm that our generalization of the treatment effects model can serve as a useful tool to uncover natural history trends in longitudinal data that are obscured by endogenous treatment.\\

\end{spacing}

\clearpage

\addtocounter{page}{-1}

\begin{spacing}{2.0}

\clearpage

\section{Introduction}

Often in epidemiologic cohort studies involving cardiovascular biomarkers, a large segment of study participants are on one or more medications intended to specifically alter those biomarker values. For such subjects, the treated biomarker value is perhaps of greatest relevance when seeking to understand subject-specific risk of subsequent cardiovascular events. However, when seeking to uncover associations between certain long-term exposures (e.g., age or gender) and a biomarker outcome, the natural history of the biomarker that would have occurred in the absence of treatment is more informative than the value under observed treatment. For treated subjects, the biomarker's natural history is contaminated by endogeneity in that participants differing in their medication use status tend to differ in their underlying biomarker values due to unmeasured confounding by indication. Since the goal of treatment is to shift biomarker values toward a healthier range, the association between an exposure of interest and the observed biomarker value cannot be expected to provide an adequate representation of the natural history association that would have occurred had treatment not been a factor.

For studies studies in which patients enter the cohort already on medication, pre-treatment values are typically unavailable; in this setting, endogeneity cannot be overcome by standard approaches such as regression adjustment or inverse probability weighting, in which it is assumed that all confounders are measured and properly accounted for. Heckman's hybrid model with structural shift (henceforth referred to as the ``treatment effects model") was proposed as a method to estimate treatment effects that are subject to unmeasured confounding \citep{Heckman78}. In prior work, we examined this approach specifically through the likelihood-based framework of \cite{Maddala83} in order to correct endogeneity bias for estimation of the natural history association between a predictor and a biomarker in cross-sectional data; from this vantage point, we view treatment as a nuisance to be accounted rather than a parameter of interest \citep{Spieker15}. Identification of the natural history association is obtained though (1) parametric specification of a joint model for the outcome and treatment, and (2) a treatment effect homogeneity assumption; these assumptions allow us to bypass both the conditional ignorability assumption typically associated with inverse probability of treatment weighting methods \citep{Robins00} and the exclusion restriction assumption typically associated with instrumental variable methods \citep{Imbens94}. Recently, we developed an extension of the treatment effects model to allow treatment effects to vary across covariates, thereby relaxing the homogeneity assumption \citep{Spieker18}. Further, our prior exploration of this approach has demonstrated that this modeling framework exhibits fairly robust behavior for estimation of the natural history association under violations to a range of assumptions, including misspecified error distributions, variable omission, and non-differential exposure mismeasurement.

Many longitudinal studies over the last two decades have sought to characterize trends in mean LDL across age and/or by gender in sub-populations of interest (\citealp{Carroll05}; \citealp{Duncan19}; \citealp{Gupta16}; \citealp{Russo15}; \citealp{Singh12}; \citealp{Zitnanova20}). Conclusions from these studies were largely derived on the basis of trends in observed LDL values, and were not based on methods that attempt to correct for endogenous medication use. Although the framework of Heckman's treatment effects model can be used to correct endogeneiety bias in the setting of cross-sectional observational data, generalizations to longitudinal studies in which observations are correlated within subjects over time remain unexplored. In this manuscript, we seek to address this methodological gap by developing and justifying a longitudinal endogeneity model (LEM).

The remainder of this manuscript is organized as follows. In Section 2, we provide background on the assumptions of Heckman's treatment effects model for estimation of natural history associations in cross-sectional data, including our extension to relax the assumption of treatment effect heterogeneity. We further provide information on computational strategies to aid the implementation of this approach. In Section 3, we present our proposed longitudinal endogeneity model, focusing specifically on the assumptions necessary for identification. In Section 4, we conduct a variety of simulation studies to evaluate the finite sample properties of the LEM. In Section 5, we illustrate the use of our approach for estimation of LDL trends by age and gender using data from the Multi-Ethnic Study of Atherosclerosis. Finally, we conclude in Section 6 with a discussion of our findings, conclusions, and possible directions for further research.

\section{Background}

In this section, we summarize how Heckman's treatment effects model can be used to achieve the goal of estimating what we refer to as the natural history association between a predictor of interest and a biomarker outcome in cross-sectional data, including our prior extension to accommodate covariate-specific treatment effects \citep{Spieker18}.

\subsection{Notation, assumptions, and definitions}

Let $i = 1, \dots, N$ index independently sampled study participants, each having observed treatment status $A_i$ (assumed binary). We follow the potential outcomes notation of \cite{Rubin05}. We let $Y_i^a$ denote the potential biomarker value that would be observed under treatment $A_i = a$, and $Y_i$ the observed biomarker outcome. Each subject has total vector of exogenous covariates $\bO_i$ having length $J$ that can be subsetted in the following useful ways: we let $\bX_i \subseteq \bO_i$ denote predictors of the outcome, $\bZ_i \subseteq \bO_i$ denote predictors of treatment, and $\bW_i \subseteq \bO_i$ denote covariates that modify the association between treatment and the outcome. Importantly, covariates appearing in $\bX_i$, $\bZ_i$, and $\bW_i$ need not be mutually exclusive. A directed acyclic graph (DAG) is depicted in Figure 1 in order to illustrate the presumed relationship between variables.

\begin{figure}[h!]
\centering
\includegraphics[width = 2.5in]{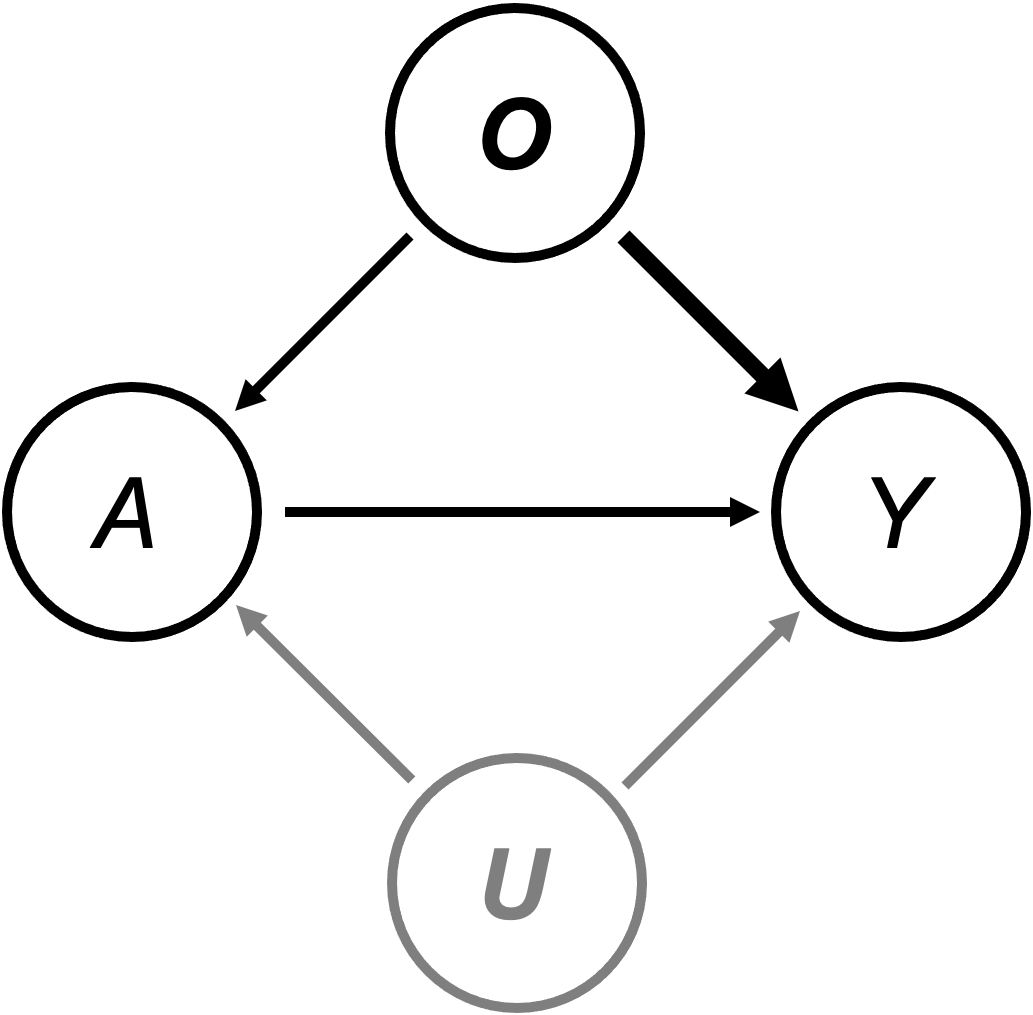}
\caption{DAG depicting the relationship between variables. The total covariate vector of exogenous covariates, $\bO$, contains variables that can be associated with either $A$ alone, $Y$ alone, or both; for ease of presentation, we do not partition $\bO$ in this graphical depiction. The parameter of interest can be described as the association between any possible subset of the covariates in $\bO$ and the off-treatment potential outcome, $Y^0$ (shown in bold). Importantly, $\bU$ contains a collection of unobserved confounders (e.g., a pre-treatment biomarker value), such that $A$ is endogenous.}
\end{figure}

Our goal is to estimate the association between (at least one predictor appearing in) $\bX$ and $Y^0$, the biomarker value that would be observed in the absence of medication use. Under a linearity assumption, a parameter of this sort can be expressed succinctly as:
\begin{eqnarray*}
\beta_j = \E[Y^0|\bX = (x_1, \dots, x_{j - 1}, x_j + 1, x_{j + 1} \dots, x_{J_X})]\\
~ & ~ & \hspace{-2.9in} - \hspace{1mm} \E[Y^0|\bX = (x_1, \dots, x_{j - 1}, x_j, x_{j + 1} \dots, x_{J_X})],
\end{eqnarray*}
where $J_X$ denotes the number of covariates appearing in $\bX$. Since $Y^0$ is only observed for those subjects who are not treated, standard regression-based approaches for estimation of $\beta_j$ are not valid. Importantly, a traditional causal interpretation for $\beta_j$ may not be appropriate, as $X_j$ need not be manipulable (e.g., age). As previously discussed in Section 1, the population average causal effect of the endogenous treatment variable, $A$, on the outcome, $Y$, is not itself of interest the setting we are describing in this work; instead, treatment serves as a nuisance in the goal of estimating $\bbeta$. Still, certain conditions must hold regarding the relationship between these variable in order for $\bbeta$ to be identified. To that end, we make the following initial identifying assumptions:
\begin{enumerate}
\item No interference: $Y_i^a \ci A_{i'}$, $1 \leq i \neq {i'} \leq N$.
\item Consistency: $Y_i = Y_i^{A_i}$.
\item Conditional homogeneity: $\E[Y^1 - Y^0|\bO] = \bW^T\boeta$.
\end{enumerate}
The assumption of no interference states that the treatment status of one individual does not influence the potential outcome of another individual. Taken together with the assumption of no interference, consistency ensures that for each subject, the observed outcome corresponds to the potential outcome under the observed treatment. The conditional homogeneity assumption is an extension of the treatment effects model that allows any arbitrary subset of observed covariates to serve as possible modifiers of the treatment effect \citep{Spieker18}. As is implied by the notation, all covariates in $\bO$ that serve as effect modifiers must be included in $\bW$.

The common assumption of conditional ignorability (namely, that $Y^a \ci A|\bO$) is not one of the identifying assumptions we will plan to invoke. Specifically, we wish to allow for the possibility of unmeasured confounding of the treatment-outcome relationship. Further, we do not propose an exclusion restriction or monotonicity assumption (Figure 1). Identification of parameters of interest will instead depend upon assumptions specific to the estimation procedure, elaborated on in Section 2.2. 

\subsection{Estimation in cross-sectional data}
For ease of notation, let $\bY = (Y_1, \dots, Y_N)^T$ denote the outcome vector, and let $\bX$, $\bZ$, and $\bW$ denote design matrices for each of the respective covariate groupings. The estimation approach described by \cite{Spieker18} builds upon the maximum likelihood estimation procedure described by \cite{Maddala83} in order to estimate parameters from the following system of simultaneous structural equations:
\begin{equation}
\begin{aligned}
\bY &= \bX\bbeta + \bW\boeta \circ \bA + \beps;\\
\bA^* &= \bZ\balpha + \bgamma.
\end{aligned}
\end{equation}
Here, $\beps$ and $\bgamma$ denote vectors of error terms that are each independent and identically distributed across subjects, but are possibly themselves correlated. Note that $A^*$ denotes a continuous latent underlying treatment variable, from which the dichotomous treatment $A = \textbf{1}(A^* > 0)$ is observed. This parameterization is equivalent to a probit model for the probability of treatment, but the error representation is particularly convenient in forming the maximum likelihood estimator. If $\btheta$ denotes all model parameters, a general form for the log-likelihood can be expressed as follows:
\begin{equation}
\begin{aligned}
\bell(\btheta; \bO, \bA, \bY) = \sum_{i = 1}^{N} \left[\log p_{\btheta}(Y_i|\bO_i) + \log \int_{\mathcal{A}_i} p_{\btheta}(A_i^*|\bO_i)dA_i^*\right],\\
\end{aligned}
\end{equation}
where $\mathcal{A}_i = (-\infty, 0)$ if $A = 0$ and $(0, \infty)$ if $A = 1$. Not all parametric specifications give rise to identifiability of $\bbeta$. One notable example in which identifiability is achieved is the setting in which the error terms are presumed to follow a common bivariate normal distribution with outcome error variance $\sigma_Y^2$ and correlation $\rho$. As the latent treatment error variance, $\sigma_A^2$ is not itself identifiable, the usual procedure it to set $\sigma_A^2 = 1$, from which weak identifiability of $\rho$ and $\balpha$ are achieved \citep{Freedman10}. Under the bivariate normal specification, the log-likelihood can be expressed for $\btheta = (\balpha, \bbeta, \boeta, \sigma_Y, \rho)$ as follows:
\begin{equation}
\begin{aligned}
\bell(\btheta; \bO, \bA, \bY) = \sum_{i = 1}^{N} \left[\log \phi\left(\frac{Y_i - \bX_i^T\bbeta - \bW_i^T\boeta \times A_i}{\sigma_Y}\right) -\log \sigma_Y \right.\\
~ & ~ & \hspace{-3.2in} \left. + \hspace{1mm} \log \Phi\left((-1)^{1 - A_i}\frac{\bZ_i^T\balpha + \rho(Y_i - \bX_i^T\bbeta - \bW_i^T\boeta \times A_i)/\sigma_Y}{\sqrt{1 - \rho^2}}\right) \right].
\end{aligned}
\end{equation}

Note that $\phi(\cdot)$ and $\Phi(\cdot)$ denote the standard normal density and cumulative distribution functions, respectively. Like $\bbeta$, $\boeta$ is identifiable, as proven in previous work \citep{Spieker18}. Further note that the parametric specification of a probit model for medication use implies the usual positivity assumption (namely, that $\Pro(A = 1|\bZ = \bz) > 0$ $\forall \bz$ such that $f_{\bZ}(\bz) > 0$). For the purposes of this work, in which our primary focus is on $\bbeta$, we will not give $\boeta$ much attention and will consider it as a nuisance parameter.

\subsection{Likelihood evaluation, software, and robustness}
We briefly comment on evaluation of the log-likelihood of Equation (3). First, we note that in order for the negative log-likelihood to possess a global maximum on the interior of the parameter space, we require the easily evaluable condition of outcome overlap, meaning that $\bigcap_{a \in \lbrace 0, 1 \rbrace} (\min_{A_i = a} Y, \max_{A_i = a} Y) \neq \emptyset$. Second, the resulting score functions do not possess a closed-form analytic solution; numerical techniques are required to evaluate the likelihood.

The correlation parameter, $\rho$, must satisfy the condition that $-1 < \rho < 1$; to that end, one may wish to re-parameterize via either the relationship $\rho = 2\arctan(\varrho)/\pi$ or $\rho = 2/(1 + \exp(-\varrho)) - 1$. Since each of these mappings is bijective, one can optimize over $\varrho$ to obtain the maximum likelihood estimate of $\rho$ in order to avoid computational challenges associated with constraining the optimization. The Broyden–Fletcher–Goldfarb–Shanno (BFGS) algorithm can be used to find a numeric solution to the likelihood \citep{Fletcher87}.

We have made an R package available CRAN titled \texttt{endogenous}. The function \texttt{hybrid} within this package can be used to maximize the likelihood of Equation (3). Parameters are initialized based on standard regression approaches, and the BFGS algorithm is used to iterate toward the maximum likelihood solution based on the gradient, included to provide more computationally efficient and algorithmically stable estimation as compared to quasi-Newton based techniques. Standard likelihood theory allows the use of the asymptotically efficient inverse Fisher information matrix for settings in which the parametric assumptions are thought to hold. The \texttt{hybrid} function also contains a robust sandwich variance estimator in the spirit of \cite{White80} to provide valid standard errors under model misspecification. Either can be used to formulate confidence intervals and conduct robust Wald-based hypothesis tests, described in further detail by \cite{Spieker18}.

Despite what are sometimes viewed as stringent parametric assumptions, prior work has demonstrated that this modeling framework exhibits fairly robust behavior for estimation of $\bbeta$ in a variety of settings, including skewed or heavy-tailed errors, various forms of variable omission, and non-differential exposure mismeasurement \citep{Spieker15}.

\section{The longitudinal endogeneity model}

In this section, we propose a longitudinal endogeneity model (LEM) to estimate marginal natural history associations in longitudinal data.

\subsection{Expanding notation, assumptions, and definitions}

We expand the notation of Section 2.1 as follows, following the potential outcomes notation of \cite{Robins86} for longitudinal data. Suppose each subject has $T$ observation times (we will later discuss variable observation times). We let $\bO_{it}$, $A_{it}$, and $Y_{it}$ denote the covariate vector, treatment status, and biomarker outcome, respectively, for subject $i$ at time $t$. We use overbar notation in defining variable history (e.g., $\overline{A}_{it} = (A_{i1}, \dots, A_{it})$, treatment history up through time $t$). Analogously, let $\underline{A}_{it} = (A_{it}, \dots, A_{iT})$ denote treatment from time $t$ onward. For convenience of notation, we let $\overline{A}_i = \overline{A}_{iT}$ denote the entire treatment history. Potential outcomes are in turn defined by hypothetical treatment history, $Y_{it}^{\overline{a}}$. We expand our goal of Section 2.1 to estimation of the population-average natural history association between $\bX$ and $Y$ that would be observed in the absence of medication use:
\begin{equation}
\begin{aligned}
\beta_j = \E[Y_t^{\overline{a} = 0}|\bX_t = (x_{t1}, \dots, x_{t(j - 1)}, x_{tj} + 1, x_{t(j + 1)} \dots, x_{tJ_X})]\\
~ & ~ & \hspace{-2.9in} - \hspace{1mm} \E[Y_t^{\overline{a} = 0}|\bX_t = (x_{t1}, \dots, x_{t(j - 1)}, x_{tj}, x_{t(j + 1)} \dots, x_{tJ_X})].
\end{aligned}
\end{equation}
We update the identifying assumptions of Section 2.1 to generalize to the longitudinal setting as follows:
\begin{enumerate}
\item No interference: $Y_{it}^{\overline{a}} \ci A_{j't'}$, $1 \leq i \neq i' \leq N$, $1 \leq t \neq t' \leq T$.
\item Consistency: $Y_{it} = Y_{it}^{A_{it}}$.
\item Conditional homogeneity: $\E[Y_t^{\overline{a}_{t - 1}, 1, \underline{a}_{t + 1}} - Y_t^{\overline{a}_{t - 1}, 0, \underline{a}_{t + 1}}|\bO_t] = \bW_{t}^T\boeta$ $\forall$ $(\overline{a}_{t - 1}, \underline{a}_{t + 1})$.
\end{enumerate}
The generalization of Assumptions 1 and 2 to the setting of repeated measures is straightforward. Conditional homogeneity implies that, conditional on $\bW_t$, the causal effect of treatment at time $t$ is homogeneous with respect to all other covariates in $\bO_t$.

\subsection{Semi-parametric estimation}
Given the joint nature of the modeling approach (i.e., between the treatment $A$ and the outcome $Y$), there are a resulting four major classes of correlation that must be considered when both $A$ and $Y$ are measured over time. Figure 2 illustrates these correlation classes in the simple setting of two observations. The first form of correlation is one we have already considered, and that is the correlation between the error terms of the treatment and outcome models at concurrent times: $\rho = \text{Corr}(\epsilon_{it}, \gamma_{it})$. The other types of correlation are longitudinal in nature, and we make no assumption about their values or functional structures.

\begin{figure}[h!]
\centering
\includegraphics[width = 5.9in]{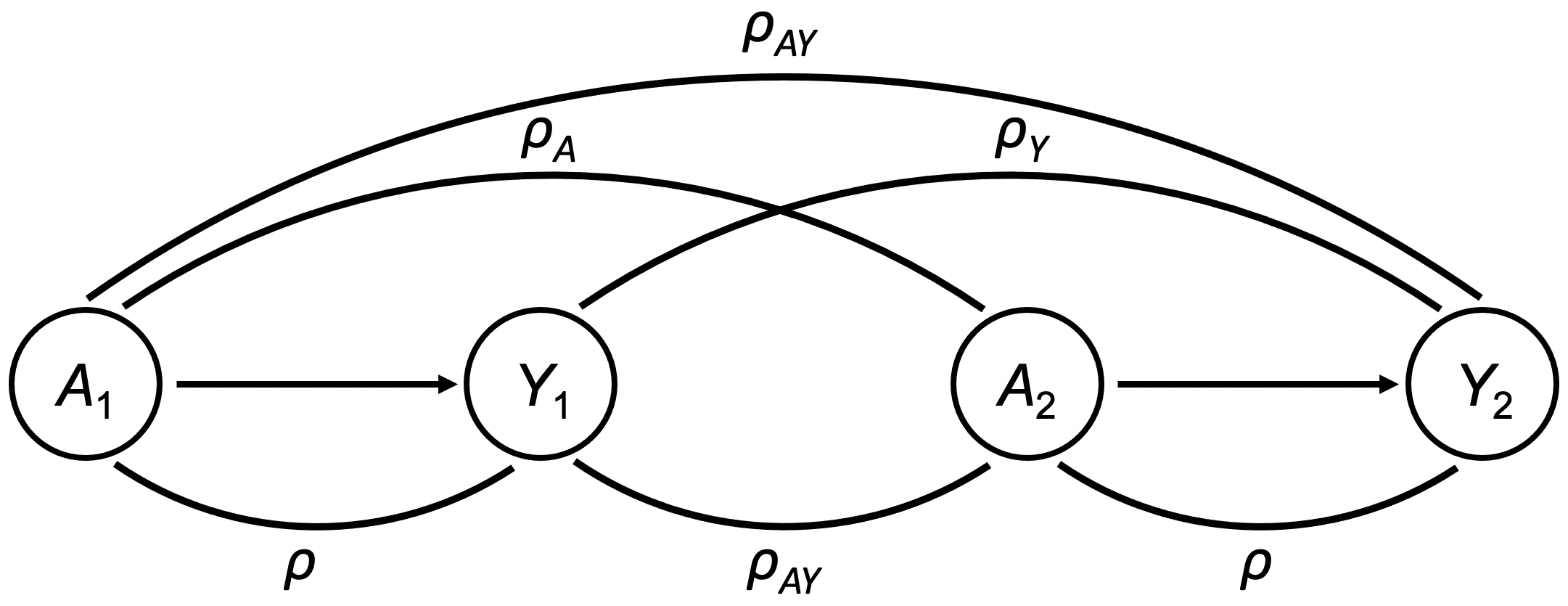}
\caption{Illustration of the different classes of correlation under consideration. Of note, $\rho_A$, $\rho_Y$, and $\rho_{AY}$ may be nonzero, though they are neither modeled nor estimated in our proposed LEM. }
\end{figure}

Let $\bell_{it}(\btheta; \bO_{it}, A_{it}, Y_{it})$ denote the contribution of subject $i$ to the log-likelihood for $\btheta$ at observation $t$. We propose the following semi-parametric estimating equations for $\btheta$ based on the total of $\sum_{i = 1}^{N} T$ observations:
\begin{equation}
\begin{aligned}
\mathcal{U}_N(\btheta) &=& \sum_{i = 1}^{N} \sum_{t = 1}^{T} \frac{\partial \bell_{it}(\btheta; \bO_{it}, A_{it}, Y_{it})}{\partial \btheta} = \textbf{0}.
\end{aligned}
\end{equation}
In contrast to the likelihood proposed for cross-sectional data, these equations are not fully parametric score equations as they do not involve the correlation between or within observations over time. Although such correlations are all but certain to be nonzero in most practical settings, they need not be estimated. As the cross-sectional score equations have expectation zero, it follows that $\E[\mathcal{U}_N(\btheta)] = \textbf{0}$, and hence these estimating equations can be solved for $\btheta$ to obtain an estimator $\widehat{\btheta}$. A robust cluster-based covariance estimator can be used to form standard errors and conduct inference on $\bbeta \subseteq \btheta$. Denoting the indices of $\bbeta$ as a subset of $\btheta$ as 1:$(J_X + 1)$ without loss of generality,
\begin{equation}
\begin{aligned}
\widehat{\Var}(\widehat{\bbeta}) &=& \left[\left[\left.\frac{\partial\mathcal{U}(\btheta)}{\partial\btheta}\right|_{\btheta = \widehat{\btheta}}\right]^{-1}\left[\sum_{i = 1}^{N} \mathcal{U}_i(\widehat{\btheta}) \mathcal{U}_i(\widehat{\btheta})^T\right]\left[\left.\frac{\partial\mathcal{U}(\btheta)}{\partial\btheta}\right|_{\btheta = \widehat{\btheta}}\right]^{-T}\right]_{[(1:(J_X + 1)), (1:(J_X + 1)]}.
\end{aligned}
\end{equation}
The \texttt{hybrid} function in the R package \texttt{endogenous} accommodates repeated measures on subjects by allowing a subject identifier argument. The robust cluster-based covariance estimator is the only available estimator in this setting, although a nonparametric bootstrap procedure could be manually programmed in its place \citep{Davison97}.

\subsection{Variable follow-up times and informative missingness}

The proposed approach can be extended to accommodate variation in the number of observations per subject. Assumptions 1 through 3 generalize in the expected way, and the inner sum of the estimating equations is taken over the number of observation times for the particular subject indexed in the outer sum. Consistency of $\widehat{\bbeta}$ for $\bbeta$ will not generally hold if cluster size is informed by the latent error terms. This is analogous to assumptions made by approaches such as generalized estimating equations (GEE) with a working independence correlation structure \citep{Hoffman01}.
 
\section{Simulation study}

We present the results of a simulation study to elucidate the finite-sample properties of the proposed semi-parametric estimation procedure of Section 3.2. In particular, we focus on bias, standard errors, and coverage probability for estimation of $\bbeta$. All simulation studies were conducted using R, version 4.0.2 (2020).

\subsection{Simulation 1: Panel data}

Consider a study of $N = 500$ independent observations, each with $T = 3$ observations. Let $\bO_{it}$ denote a total covariate vector of length $J = 7$, with $\bX_{it} = (1, O_{it1}, O_{it4}, O_{it5}, O_{it7})$, $\bZ_{it} = (1, O_{it2}, O_{it4}, O_{it6}, O_{it7})$, and $\bW_{it} = (1, O_{it3}, O_{it5}, O_{it6}, O_{it7})$; that is, there are seven possible covariates, withcombinations of four related to each of the outcome, treatment, and effect size. The purpose of this formulation is so that there is a predictor appearing in each of the three subsets alone, a predictor appearing in each of the three possible pairs of subsets, and a predictor that appears in all three; of note, we are also allowing an intercept in this formulation. We generate $\bO_i$ from a multivariate normal distribution with the following correlations: (1) a common correlation of $0.20$ between covariates measured at the same time, (2) a common correlation of $0.30$ within each variable over time, and (3) a common correlation of $0.10$ between different covariates measured at different times. Further, suppose the treatments $A_{it}$ and outcomes $Y_{it}$ are generated according to the following mechanism: 
\begin{equation}
\begin{aligned}
\begin{bmatrix} \bY_{i} \\ \bA_{i}^* \end{bmatrix} = \begin{bmatrix} \bX_i\bbeta + \bW_i\boeta \\ \bZ_i\balpha \end{bmatrix} +
\begin{bmatrix} \beps_i \\ \bgamma_i \end{bmatrix}; \hspace{0.1in} \begin{bmatrix} \beps_i \\ \bgamma_i \end{bmatrix} \sim \mathcal{N}\left(\textbf{0}, \begin{bmatrix} \bSigma_{11} & \bSigma_{12} \\ \bSigma_{12}^T & \bSigma_{22}\end{bmatrix} \right).
\end{aligned}
\end{equation}
Here, $\bSigma_{11}$ is an exchangeable covariance matrix having outcome error variance of $\sigma_Y^2 = 1.0$ and a correlation of $\rho_Y = 0.60$ between outcome errors over time; $\bSigma_{12}$ includes a correlation of $\rho = 0.50$ between the latent treatment error $\gamma_{it}$ and the concurrent outcome error $\epsilon_{it}$ on the diagonal entries and a correlation of $\rho_{AY} = 0.20$ between the two errors at different times on off-diagonal entries; $\bSigma_{22}$ contains diagonal entries of $\sigma_A^2 = 1$ for the treatment error variance, off-diagonal entries of $\rho_{A} = 0.50$, signifying the correlation in treatment errors over time. We set $\bbeta = \balpha = (0, 1, 1, 1, 1)^T$, and we set $\boeta = (0, 0.20, 0.20, 0.20, 0.20)^T$ in this simulation.

As a comparator method, we use GEE with a working independence correlation structure, adjusted for treatment. We conduct one-thousand simulation replicates under this setup, extracting at each replicate the corresponding point estimates, robust cluster-based standard errors, and indicator of coverage based on symmetric 95\% Wald-based confidence intervals. Results are presented in Table 1.

\begin{table}[h!]
\caption{Simulation study results for panel data (i.e., constant cluster size). Presented are the mean estimates across simulation replicates, the empirical standard error (ESE), the average estimated standard error ($\widehat{\text{SE}}$), and the coverage probability (CP) for each of $\beta_0$ through $\beta_4$.}
\centering
\begin{tabular}{rccccccccc}
~ & \multicolumn{4}{c}{LEM} & & \multicolumn{4}{c}{GEE}\\ \cmidrule{2-5}\cmidrule{7-10}
Coefficient & Estimate & ESE & $\widehat{\text{SE}}$ & CP & & Estimate & ESE & $\widehat{\text{SE}}$ & CP \\ \hline
$\beta_0 = 0$  & 0.00 & 0.067 & 0.065 & 0.94   && -0.40 & 0.056 & 0.055 &  0.00 \\
$\beta_1 = 1$   & 1.00 & 0.030 & 0.030  & 0.95   &&   1.00 & 0.031 & 0.031 & 0.96   \\
$\beta_2 = 1$  & 1.00 & 0.036 & 0.036  & 0.94   &&   0.92 & 0.034 & 0.034 & 0.38  \\
$\beta_3 = 1$  & 1.00 & 0.041 & 0.041   & 0.95   &&   0.90 & 0.032 & 0.032 & 0.11   \\
$\beta_4 = 1$  & 1.00 & 0.048 & 0.047  & 0.95   &&   0.82 & 0.035 & 0.034 & 0.00  \\ \hline
\end{tabular}
\end{table}

Based on the results of this study, we see that the LEM achieves low bias under this setup, with standard errors that, on average, represent the true repeat-sample variability of the estimates across Monte Carlo iterations, and coverage that is close to the desired 95\%. Estimates produced by the GEE approach are not, in general, unbiased for $\bbeta$, although the robust standard errors appear to capture the repeat-sample variability well. We make note that GEE does achieve low bias for estimation of $\beta_1$. Recall that the variable $O_1$ is associated with neither the treatment assignment nor the effect of the treatment, and so endogeneity does not pose a challenge to estimation of $\beta_1$ using standard approaches. The relative loss of efficiency associated with the LEM is not surprising given that it involves estimation of a substantially larger number of parameters.

\subsection{Simulation 2: Completely random cluster size}

We conduct a simulation study under the same general setup as that of Section 4.1, with completely random variation in subject-specific observation times. In particular, we introduced a $1/3$ probability that a subject's second observation was missing, and a $1/2$ probability a subject's third observation was missing. Conclusions from this simulation study are analogous to those of Section 4.1, and so we do not supply them in detail.

\subsection{Simulation 3: Exposure-dependent cluster size}

We conduct a simulation study under the same general setup as that of Section 4.1, with another example of variable observation times, $T_i$. The indicator of missingness for subject $i$ at time $t$ was generated to depend upon covariates as follows:
\begin{equation}
\begin{aligned}
M_{it} \sim \text{Bernoulli}\left(p = \text{expit}\left(-1 + 0.2\times\sum_{j = 1}^{7} O_{itj}\right)\right).
\end{aligned}
\end{equation}

This results in an overall missingness rate of approximately 30\%. As expected, conclusions from this simulation study are analogous to those of Section 4.1, and so we do not supply them in detail.

\subsection{Simulation 4: Outcome-dependent cluster size}

Of interest is to evaluate the performance of the longitudinal endogeneity model under a missingness mechanism that depends upon the outcome. We conduct a simulation study under the same general setup as that of Section 4.1, with the indicator of missingness or subject $i$ at time $t$ generated as:
\begin{equation}
\begin{aligned}
M_{it} \sim \text{Bernoulli}(p = 0.1\times \textbf{1}(Y_{it} \leq -1) + 0.4\times\textbf{1}(-1 < Y_{it} \leq 2) + 0.7\times\textbf{1}(Y_{it} \leq 2)).
\end{aligned}
\end{equation}

Results from this simulation study are depicted in Table 2. As expected, the LEM produces estimates that are biased; in settings of informative cluster size, the LEM cannot be expected to produce consistent estimates of $\bbeta$. 

\begin{table}[h!]
\caption{Simulation study results for panel data (i.e., constant cluster size). Presented are the mean estimates across simulation replicates, the empirical standard error (ESE), the average estimated standard error ($\widehat{\text{SE}}$), and the coverage probability (CP) for each of $\beta_0$ through $\beta_4$.}
\centering
\begin{tabular}{rccccccccc}
~ & \multicolumn{4}{c}{LEM} & & \multicolumn{4}{c}{GEE}\\ \cmidrule{2-5}\cmidrule{7-10}
Coefficient & Estimate & ESE & $\widehat{\text{SE}}$ & CP & & Estimate & ESE & $\widehat{\text{SE}}$ & CP \\ \hline
$\beta_0 = 0$  & -0.14 & 0.076 & 0.075 & 0.55   && -0.51 & 0.066 & 0.064 &  0.00 \\
$\beta_1 = 1$   & 0.98 & 0.036 & 0.036  & 0.90   &&   0.97 & 0.037 & 0.037 & 0.89   \\
$\beta_2 = 1$  & 0.98 & 0.045 & 0.043  & 0.90   &&   0.91 & 0.042 & 0.040 & 0.36  \\
$\beta_3 = 1$  & 0.97 & 0.047 & 0.046   & 0.88   &&   0.89 & 0.038 & 0.038 & 0.18   \\
$\beta_4 = 1$  & 0.97 & 0.051 & 0.052  & 0.91   &&   0.82 & 0.041 & 0.041 & 0.01  \\ \hline
\end{tabular}
\end{table}

\section{Age and LDL: The Multi-Ethnic Study of Atherosclerosis}

The Multi-Ethnic Study of Atherosclerosis (MESA) is a longitudinal cohort study of 6,814 men and women from six U.S. communities. At the time of first observation, participants ranged from 45 to 84 years of age. The demographic breakdown of the study is as follows: 47\% male, 38\% white, 28\% African-American, 22\% Hispanic, and 12\% Chinese-American. This study was designed to provide insights into the prevalence and progression of subclinical cardiovascular disease. We use data from the exams of MESA five exams, collected over ten years. All subjects provided written informed consent. Further details regarding sampling, recruitment, and data collection are reported elsewhere \citep{Bild02}. Of note, 16.1\% of study participants reported use of at least one lipid-lowering drug at the time of their first observation. The study prevalence of lipid-lowering medication use increased during the period of time over which subjects were observed, with 39.0\% of the 4,464 subjects observed for a fifth visit having reported use of at least one lipid-lowering drug at the time of their fifth observation. Participants were aged 45 to 84 years at the baseline observation; by the time of the fifth observation, the maximum age was 94 years.
\begin{figure}[h!]
\centering
\includegraphics[width = 3.0in]{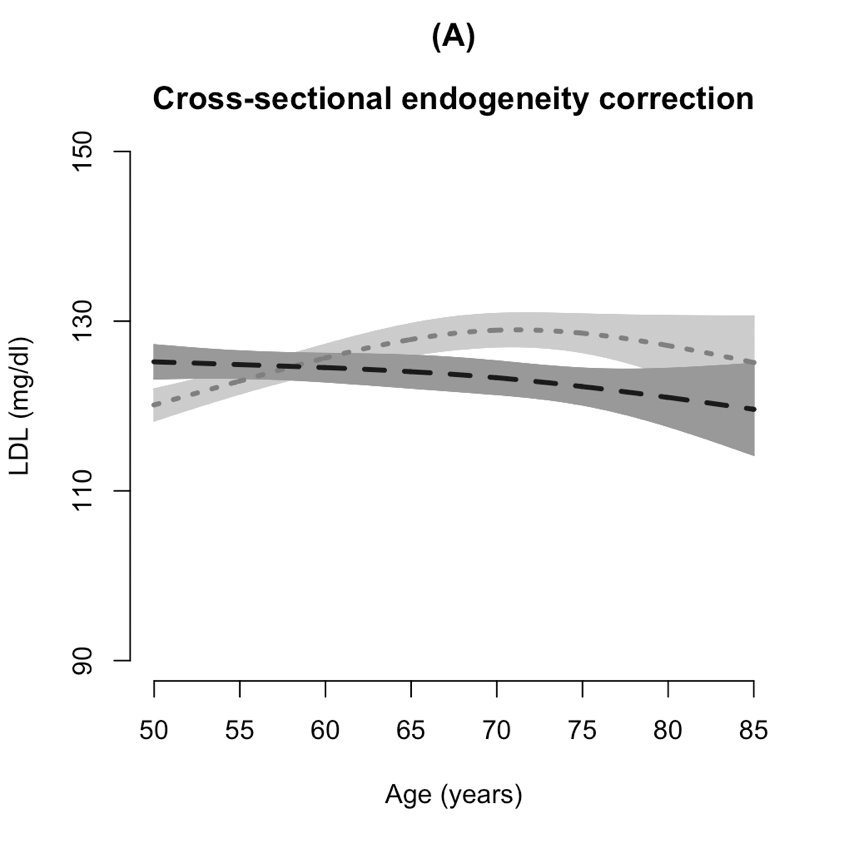} \includegraphics[width = 3.0in]{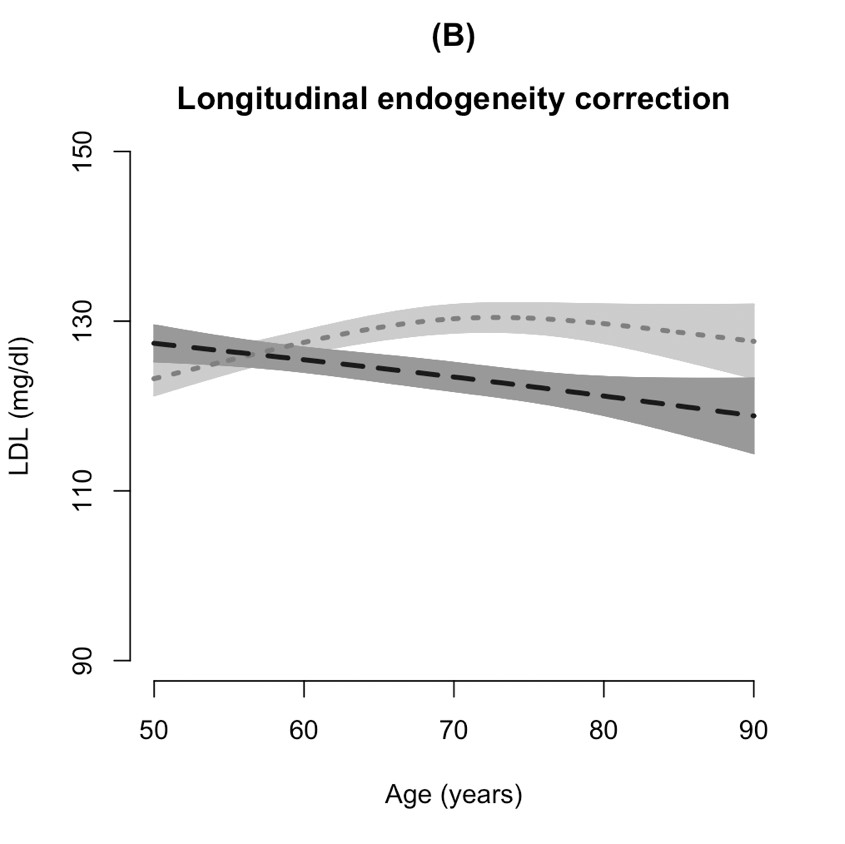}\\
\includegraphics[width = 3.0in]{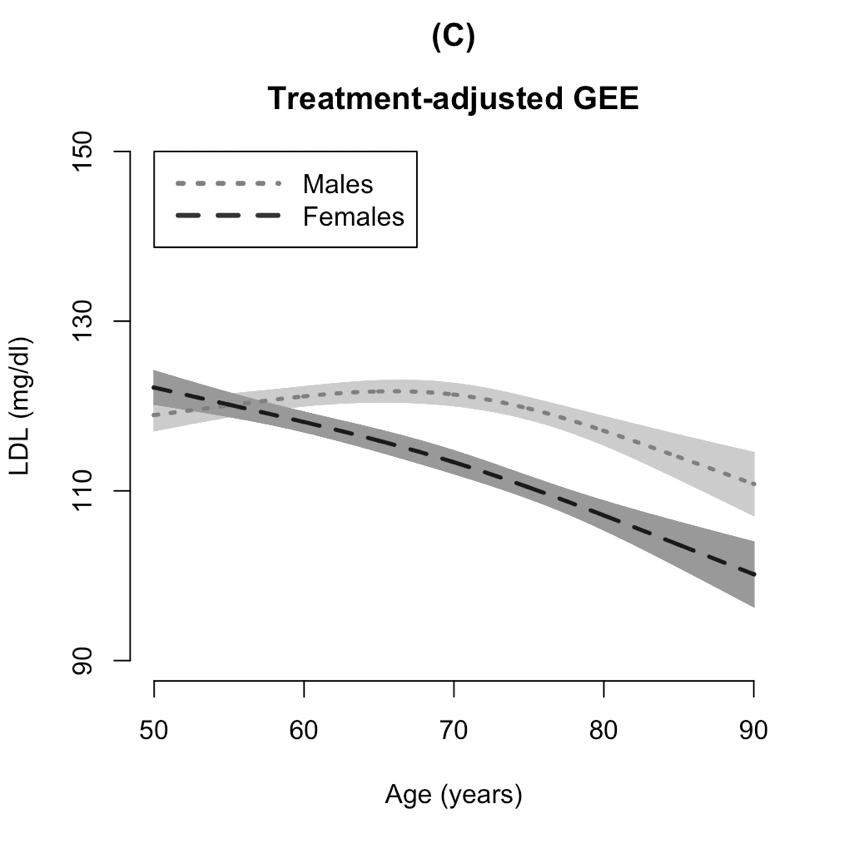} \includegraphics[width = 3.0in]{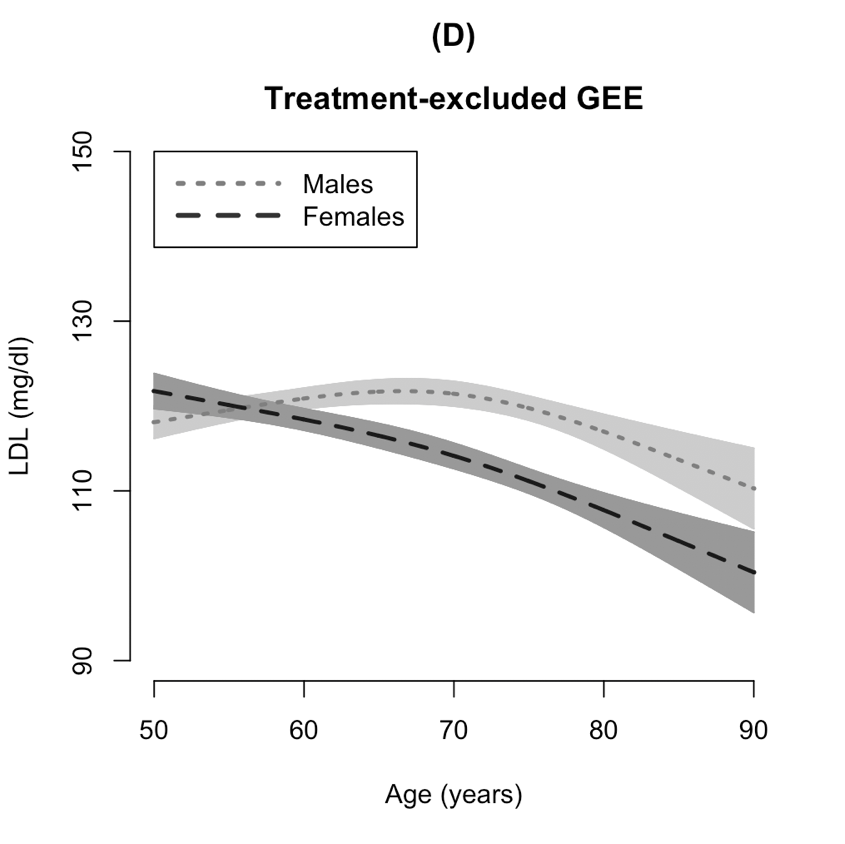} 
\caption{Predicted mean LDL across levels of age, stratified by gender, for each of four methods: (A) including the endogeneity correction in the cross-sectional data, depicted over the range of baseline age; (B) longitudinal endogeneity model; (C) treatment-adjusted GEE; (D) treatment-excluded GE. The shaded bands included represent point-wise 95\% confidence intervals.}
\end{figure}

We use these data to illustrate the utility of the LEM in order to estimate and compare natural history age trends in LDL by gender. Treatment was considered as the use of any lipid-lowering medications. In the outcome model of the LEM, we included gender, a natural cubic spline for age with knots at 55, 70, and 85 years, and an interaction between the two. The medication use (probit) model included age category, gender, race category, diabetes status, health insurance status, and the Framingham risk score. Treatment effect magnitude was permitted to vary with continuous age, gender, and lipid drug type, categorized as follows: 0 = statin; 1 = other (e.g., fibrates, niacin, bile acid resins).

We include results from three comparator models in this illustration. In the first, we apply the treatment effects model (with covariate-specific treatment effects) to data from the first observation---i.e., so that we may apply the treatment effects model suitable for cross-sectional data. We further estimate the parameters of two separate GEE models with a working independence correlation structure: one that adjusts for medication use (0 = none; 1 = statins, 2 = other), and one that excludes all observations in which a subject was treated at the time of measurement.

Figure 3 presents the predicted age trends by gender and corresponding point-wise 95\% confidence intervals for each of the four approaches. In comparing the results of the cross-sectional and longitudinal models, we note little difference in the overall age-trends by gender. Unsurprisingly, the confidence bands for the LEM are narrower as compared to the cross-sectional counterpart. This is consistent with the idea that both models are targeting the same parameter (namely the natural history trend that would be observed in the absence of medication use). The LEM has an advantage over the cross-sectional method in being able to estimate the natural history trend more efficiently by making use of additional observations, as evidenced by narrower confidence bands. That the cross-sectional model displays very wide confidence intervals for larger values of age is a product of the fact that there were fewer data points at older ages; of note, we only show the predicted results through 84 years in the cross-sectional model, the maximum baseline age. Each model correcting for endogeneity displays graphical evidence of a decrease in off-treatment LDL at older ages. This trend would be consistent with poorer liver function associated with lower age.

The GEE models represent trends in the observed biomarker. The overall trends based on the GEE model suggest a much sharper downward trend as compared to the models that account for endogeneity. In the exclusion-based model, this is easily explained by a problem of selection bias, in which those with a higher underlying LDL value are systematically excluded from the analysis. That treatment adjustment does little to correct this is attributable to endogeneity of treatment; the adjusted model conditions on observed treatment status but does not handle the correlation between treatment and the error term. The GEE models also tend to suggest much stronger evidence of a difference in mean observed LDL between genders as compared to the models for the natural history trend. Interestingly, this result is consistent with a prior study that concluded that mean LDL shows substantives decreases with age \cite{Ferrara97}, in which the study investigators adjusted for medication use in both cross-sectional and longitudinal regression models.

\section{Discussion}

In this manuscript, we have derived and presented an approach for estimation of trends in the natural history of a biomarker. Endogenous medication use acts as a contaminant when seeking to estimate marginal associations between predictors of interest and biomarkers in longitudinal data. A subject's natural history is distorted by the effects of the medication on the biomarker. Since medication users differ from non-users in their expected underlying off-medication biomarker values, and since the predictor of interest is often associated with higher medication user prevalence, na\"{i}ve approaches to account for this distortion are not appropriate. Utilizing a working independence model, as we have proposed, is a means of extending the cross-sectional treatment effects model to accommodate clustering while bypassing computational difficulties associated with full specification of a longitudinal covariance matrix. In our application, we were able to demonstrate substantial efficiency gains by making use of repeated measures on subjects.

The working independence approach as proposed in our work offers an additional advantage over full covariance specification. In particular, non-independence approaches are known to suffer from the challenge of requiring the ``full-covariate conditional mean" assumption to hold, in which the mean model for an outcome at a particular time must hold conditional on the entire covariate history rather than merely the concurrent covariates (\citealp{Diggle02}; \citealp{Pepe94}). Although this assumption may hold for certain primordial exposures that are time-stable and deterministic in nature, factors that influence medication use are typically iterative and time-dependent in nature, such that this assumption could not be guaranteed. Of note, our proposed methodology can be employed when seeking to evaluate associations between biomarkers and predictors such as race/ethnicity, genetic exposures, or chronic illnesses.

We point out that we are not offering a blanket criticism of GEE for estimation of associations. Our purpose in using GEE as a comparator method in simulations and in our application to MESA was to underscore the idea that simple regression models, commonly applied in this setting, do not provide consistent estimates of the natural history trend in the specific setting of endogenous medication use.

There are several possible directions for future work. For settings in which cluster size may be informative, \cite{Hoffman01} propose a within-cluster resampling approach in GEE models; it would be of interest to evaluate whether this approach can be generalized to the LEM in order to reduce bias arising from informative missingness. Some recent studies of biomarker trends have focused on latent class growth curves via finite mixture modeling (\citealp{Loucks11}; \citealp{deGroot14}; \citealp{Allen14}). In settings of endogenous treatment, it is all but certain that finite mixture modeling approaches will fail to capture latent underlying natural history curves; therefore, a generalization of our framework to this setting could also be of interest.
\end{spacing}

\clearpage

\noindent \textbf{Acknowledgements}

\setlength{\parindent}{0.7cm} 

{
	
	\singlespacing
	
	\noindent This work was supported by R01-HL-103729-01A1. MESA is conducted and supported by the National Heart, Lung, and Blood Institute (NHLBI) in collaboration with MESA investigators. Support for MESA is provided by contracts N01-HC- 95159, N01-HC-95160, N01-HC-95161, N01-HC-95162, N01-HC-95163, N01-HC-95164, N01-HC-95165, N01-HC- 95166, N01-HC-95167, N01-HC-95168, N01-HC-95169 and CTSA UL1-RR-024156. The authors thank the other investigators, the staff, and MESA participants for their valuable contributions. A full list of participating MESA investigators and institutions can be found at http://www.mesa-nhlbi.org.
	
}

\clearpage

{\onehalfspacing

\noindent \bibliography{arXiv}

}

 \end{document}